\documentclass[preprint,showpacs,preprintnumbers,amsmath,amssymb, superscriptaddress,longbibliography,nofootinbib]{revtex4-1}
\usepackage{graphicx} 
\usepackage{xcolor}
\usepackage{amsmath}
 \usepackage{hyperref}
\usepackage[top = 2cm, bottom = 2cm, right = 2cm, left =2cm]{geometry}
\usepackage{enumitem}
\usepackage{subcaption}
\usepackage{caption}
\begin{document}

\title{Black hole solutions with a linear equation of state in Hořava gravity and Einstein--{\ae}ther theory}

\author{Milko Estrada }
\email{milko.estrada@gmail.com}
\affiliation{Departamento de Física, Facultad de Ciencias, Universidad de Tarapacá, Casilla 7-D, Arica, Chile}

\date{\today}

\begin{abstract}
We provide a procedure to obtain black hole (BH) solutions in Ho\v{r}ava gravity and Einstein--{\ae}ther theory (HG--EA) for the spherically symmetric (SS) case with a static {\ae}ther. This procedure consists of first specifying the form of the equation of state (EoS), rather than prescribing an energy density profile. The usual EoS for the static and SS case, $\rho = -p_r$, is no longer satisfied due to the presence of the HG--EA terms. We study three linear EoS associated with: an analogue charged BH, a non-trivial extremal BH, and an ultra-relativistic stiff fluid, respectively. The HG--EA terms lead to exotic behaviors, both in the physical properties of the solutions and in their thermodynamics. In Case I, the matter sources can be interpreted as an exotic anisotropic matter distribution, giving rise to an effective electric-potential term in the geometry. In Case II, we obtain a non-trivial extremal BH solution for which the event horizon is $n_{\text{odd}}$-fold degenerate. In Case III, we find a solution with a non-trivial repulsive potential, where the influence of the HG--EA terms at short scales leads to the formation of a BH remnant whose horizon encloses a central singularity (instead of a de Sitter core as occurs in regular BHs).

\end{abstract}

\maketitle

\section{Introduction}

The observation of gravitational waves \cite{LIGOScientific:2016aoc,LIGOScientific:2017ycc} has provided strong additional support for General Relativity (GR), extending its remarkable success beyond the scale of the Solar System \cite{Damour:2007gr}. However, unresolved issues of GR at quantum scales, together with the mysterious origins of dark energy and dark matter, have motivated researchers to explore alternative or modified theories of gravity. However, applying quantum field theory (QFT) within the framework of GR to obtain a theory of quantum gravity (QG) results in a perturbatively non-renormalizable theory. A common strategy to cure ultraviolet (UV) divergences, which in some cases leads to renormalizable theories, has been to include higher-order derivative corrections of the metric in the Einstein--Hilbert action. However, this approach sometimes results in the emergence of massive gravitons in the form of ghosts, i.e., modes with negative kinetic energy. As noted in reference \cite{Wang:2017brl}, the presence of such ghosts is closely related to the fact that the modified theory contains time derivatives of order higher than two. In this regard, Ostrogradsky’s theorem states that a system is not kinematically stable if it is described by a non-degenerate Lagrangian involving higher-order time derivatives. Consequently, any higher-derivative gravitational theory that preserves Lorentz invariance (LI) and satisfies the non-degeneracy condition is inherently unstable.

Regarding LI, it is worth noting that there are observational constraints that make it phenomenologically difficult to violate this symmetry in the matter sector \cite{Liberati:2009pf}. On the other hand, in the gravitational sector, where the coupling is much weaker, such constraints are generally less stringent. Along these lines, any theory that violates Lorentz symmetry must be regarded as an effective theory in the low-energy limit. To break Lorentz symmetry while remaining explicitly diffeomorphism-invariant, as in GR, the theory must include, in addition to the metric tensor, a dynamical field capable of defining a preferred reference frame at the level of its solutions. A typical example is a unit timelike vector field, which breaks local boost (momentum) invariance but preserves local rotational symmetry. The most general theory that can be constructed by coupling such a vector field to GR, up to second order in derivatives, is known as Einstein--{\ae}ther theory \cite{Jacobson:2000xp,Jacobson:2007veq}. The vector field itself is referred to as the {\ae}ther. On the other hand, the Einstein--{\ae}ther theory, viewed as an effective field theory at low energies, can be regarded as a description of Lorentz-violating (LV) effects that might arise from a more fundamental theory of quantum gravity \cite{Barausse:2011pu}. In this way, and in relation to the aspects described in the previous paragraphs, the study of black hole solutions in Einstein--{\ae}ther gravity has attracted considerable attention in recent years \cite{Mukherjee:2024hht, Mukohyama:2024vsn, Franzin:2023rdl,Wang:2022yvi,Chan:2021ela,Adam:2021vsk}.

Another proposal for a Lorentz-violating theory of gravity is Ho\v{r}ava--Lifshitz (HL) gravity \cite{Horava:2009uw}. Some authors have suggested that this theory could provide a possible ultraviolet completion of GR. HL gravity aims to be renormalizable while simultaneously avoiding the emergence of ghosts. Specifically, its strategy involves breaking Lorentz invariance in the ultraviolet regime and including higher-order spatial derivative terms in the Lagrangian, while keeping time derivatives up to second order. In HL gravity, this approach implies the existence of a preferred spatial foliation of spacetime, which is described by a scalar field.

As noted in \cite{Barausse:2011pu}, since Einstein--{\ae}ther theory (EA theory) is a fairly general effective theory of Lorentz-violating gravity with a single preferred local timelike direction, it is reasonable to expect that the low-energy limit of Ho\v{r}ava--Lifshitz gravity bears some resemblance to it. In this context, Ref. \cite{Jacobson:2010mx} shows that, in the limit where higher-order operators beyond second order can be neglected, HL gravity is equivalent to EA theory with the additional requirement that the {\ae}ther be orthogonal to hypersurfaces at the level of the action. Moreover, \cite{Barausse:2011pu} also emphasizes that one of the connections between the two theories lies in the analogous form of their spherically symmetric solutions. This is because all spherically symmetric {\ae}ther fields are orthogonal to hypersurfaces. Consequently, all spherically symmetric solutions of Einstein--{\ae}ther theory are also solutions in the infrared limit of Ho\v{r}ava--Lifshitz gravity. However, it is worth noting that the reverse argument holds only for solutions with a regular center \cite{Blas:2010hb}. Nevertheless, without this last condition, additional solutions may exist in HL gravity.

In line with this work, as indicated in Ref.~\cite{Vernieri:2017dvi}, Ho\v{r}ava gravity admits a covariant formulation that coincides with Einstein--{\ae}ther theory when the {\ae}ther is chosen to be hypersurface-orthogonal at the level of the action. In this reference, the authors focus, for simplicity, on the case of a static {\ae}ther, finding a class of potentially viable interior stellar solutions that exhibit very rich phenomenology. Within the context of the covariant formulation of Ho\v{r}ava gravity, they propose a simple reconstruction method capable of generating anisotropic solutions. Consequently, they provide some exact, static, and spherically symmetric interior solutions of the low-energy limit of the covariantized version of Ho\v{r}ava gravity in the presence of an anisotropic fluid. In a subsequent work \cite{Vernieri:2018sxd}, the same authors, following the methodology outlined in Ref. \cite{Vernieri:2017dvi}, study exact, analytical, spherically symmetric stellar interior solutions in Ho\v{r}ava gravity and Einstein--{\ae}ther theory, considering anisotropic fluids. In summary, both references propose a specific geometric ansatz from which expressions for the energy density and anisotropic pressures in the energy-momentum tensor are obtained. In particular, in \cite{Vernieri:2018sxd}, the energy-momentum components follow the structure of relativistic polytropic equations of state. See also \cite{Vernieri:2019vlh}.

On the other hand, it is well known that one way to obtain the geometric structure of spacetime, as well as the radial and/or temporal evolution of the matter components, is by knowing the form of the equation of state (EoS). Reference \cite{Paul:2017ztx} claims that the exact form of the EoS describing the evolution of the universe is not known and therefore deriving the correct cosmological evolution remains a challenge for modern cosmology. It also states that, in general, the EoS for an anisotropic fluid adopts the general form $f(p_r, p_t, \rho) = 0$. Furthermore, it affirms that assuming a very simple formulation for the EoS makes it possible to derive the evolutionary parameters and thus compare them with observational data. In this way, linear equations of state have drawn attention in recent years both for their simplicity and their ability to represent some physically relevant models. Some examples of the use of linear EoS include: the general scenario of our universe in which its geometry is characterized by a Finslerian structure \cite{Paul:2017ztx}; the construction of compact stellar object solutions in Refs.~\cite{Thomas:2016zak,Govender:2015}; and the analysis of the evolution of gravitational collapse, which can lead either to the formation of a black hole or to a naked singularity~\cite{Goswami:2004kq,Sarwe:2012wt}.

It is worth mentioning that it is well known that various black hole solutions are supported by matter sources in the energy--momentum tensor. The main strategy for constructing such solutions has been to prescribe energy density profiles based on certain physical arguments, from which the pressure components are then obtained. Well-known examples include the energy density profiles of regular black holes, see for instance \cite{Dymnikova:1992ux}, those of black holes with an integrable singularity \cite{Estrada:2023dcj}, black holes with a cosmic-void density profile \cite{Lustosa:2025mxr}, black holes with a dark-matter density profile \cite{Xu:2018wow}, etc. It is worth mentioning that a static and spherically symmetric geometry where $g_{tt} = -g_{rr}^{-1}$ implicitly yields an equation of state of the form $\rho = -p_{r}$ in General Relativity. The presence of matter sources in the energy--momentum tensor naturally leads one to consider applying the strategy of constructing black hole solutions starting from an equation of state. This strategy has been less used than the former. However, some examples can be found in Ref. \cite{Ovalle:2018umz}, where isotropic, traceless, barotropic, and linear EoS are employed to construct black hole solutions.

In this work, motivated by the ideas discussed above, we test some linear equations of state that lead to black hole solutions in Ho\v{r}ava gravity and Einstein--{\ae}ther theory. We follow the procedure developed in Refs. \cite{Vernieri:2017dvi,Vernieri:2018sxd}, assuming a static {\ae}ther configuration. We interpret how the chosen equations of state provide new interpretations of the nature of the matter sources in this modified theory of gravity. We will analyze the influence of the Ho\v{r}ava terms on the structure of the solutions and the way in which these terms lead to exotic behaviors (when compared with General Relativity), both in the physical properties of the solutions and in their thermodynamics. Moreover, in our final case study, we will discuss how the Ho\v{r}ava terms affect the behavior at small scales and the consequences of this for the radial evaporation process. As discussed in the sixth paragraph of this Introduction, linear EoS can describe several physically relevant scenarios. In this context, we explore the role that such EoS may play in HG--EA. Since the field equations do not lead directly to exact analytical solutions for all physical situations, the strategy adopted in this work is to seek linear EoS that allow us to construct exact analytical black hole solutions in a direct manner, without resorting to more sophisticated analytical methods or numerical procedures. We then investigate the physically relevant scenario associated with each equation of state, together with the geometrical and physical properties of the resulting spacetimes. On the other hand, it is well known that black holes emit thermal radiation due to quantum effects near the event horizon \cite{Hawking:1975vcx}. We also investigate some thermodynamic properties of the resulting black hole solutions. In particular, the Hawking temperature allows us to analyze how the different linear EoS influence the thermal behavior of these solutions and the possible existence of extremal black holes, thereby providing a complementary physical characterization of the corresponding geometries and allowing us to distinguish between the different scenarios considered. For our analysis, we adopt the metric signature $(+,-,-,-)$ and perform all computations in natural geometric units, where the Newtonian gravitational constant $G$, the speed of light in vacuum $c$, and the reduced Planck constant $\hbar$ are normalized to unity, namely $G=c=\hbar=1$.

\section{A brief revision of the Covariantized version of Hořava theory for the case of spherical symmetry and static {\ae}ther}

In this section, we follow the procedure proposed in References \cite{Vernieri:2017dvi,Vernieri:2018sxd}. In the low-energy regime, the action of Hořava gravity can be written as:

\begin{equation} \label{AccionHorava}
S_H = \frac{1}{16\pi G_H} 
\int dT\, d^3x\, \sqrt{-g} \,
\left( K_{ij}K^{ij} - \lambda K^2 + \xi R + \eta\, a_i a^i \right)
+ S_m[g_{\mu\nu}, \psi],
\end{equation}
where $G_H$ denotes the effective gravitational constant, $T$ represents the preferred time coordinate, and $g$ is the determinant of the four-dimensional metric $g_{\mu\nu}$. The quantity $R$ corresponds to the Ricci scalar of the spatial hypersurfaces at constant $T$, $K_{ij}$ is the extrinsic curvature tensor with trace $K$, and $a_i = \partial_i \ln \mathcal{N}$, with $\mathcal{N}$ being the lapse function in the ADM decomposition. The term $S_m[g_{\mu\nu}, \psi]$ represents the action for the matter fields $\psi$. The constants $\{\lambda, \xi, \eta\}$ are dimensionless coupling parameters. In the limit where Hořava gravity reduces to General Relativity (GR), they take the values $\{1, 1, 0\}$. As mentioned earlier, in the specific case of a spherically symmetric spacetime with a static {\ae}ther, the authors of Refs. \cite{Vernieri:2017dvi,Vernieri:2018sxd} emphasize that the equations of motion derived from the preceding action are identical to those obtained in the Einstein--{\ae}ther framework. Therefore, in the following, we focus on the covariantized formulation of the low-energy limit of Hořava gravity, commonly known as the khronometric model. In this context, these references consider the action corresponding to the Einstein--{\ae}ther theory:

\begin{equation}
S_{\ae} = -\frac{1}{16\pi G_{\ae}} 
\int d^4x\, \sqrt{-g}\, (R + \mathcal{L}_{\ae}) + S_m[g_{\mu\nu}, \psi],
\label{eq:action_ae}
\end{equation}
where $G_{\ae}$ denotes the ``bare'' gravitational constant, and the term $\mathcal{L}_{\ae}$ is defined as:
\begin{align}
\mathcal{L}_{\ae} =&
c_1 \nabla^{\alpha}u^{\beta}\nabla_{\alpha}u_{\beta}
+ c_2 \nabla_{\alpha}u^{\alpha}\nabla_{\beta}u^{\beta} \nonumber \\
&+ c_3 \nabla_{\alpha}u^{\beta}\nabla_{\beta}u^{\alpha}
+ c_4 u^{\alpha}u^{\beta}\nabla_{\alpha}u_{\nu}\nabla_{\beta}u^{\nu}.
\label{eq:lagrangian_ae}
\end{align}
where the coefficients $c_i$ are arbitrary dimensionless constants and 
$u^{\mu}$ is a unit timelike vector field satisfying $g_{\mu\nu}u^{\mu}u^{\nu} = 1$, 
commonly referred to as the {\ae}ther field. 
To establish the connection between Hořava gravity and the Einstein--{\ae}ther theory, 
we assume that the {\ae}ther is hypersurface-orthogonal already at the level of the action, 
which locally corresponds to defining
\begin{equation} \label{aether}
u_{\mu} = \frac{\partial_{\mu} T}
{\sqrt{g^{\alpha\beta}\,\partial_{\alpha}T\,\partial_{\beta}T}},
\end{equation}

In the covariant formulation, the preferred time $T$ is represented as a scalar field, commonly referred to as the \emph{khronon}, which establishes the preferred foliation of spacetime. In our case, following Ref.~\cite{Jacobson:2010mx}, the level surfaces $T=\mathrm{const.}$ define the preferred foliation. On the other hand, following Ref.~\cite{Barausse:2011pu}, one can choose $T$ itself to be the time coordinate $t$, such that in Eq.~(4) one has $\partial_\mu T=\delta_\mu^T$. Consequently, the hypersurfaces $T=\mathrm{const.}$ coincide with the hypersurfaces $t=\mathrm{const.}$. Thus, following Ref.~\cite{Barausse:2011pu}, Eq. \eqref{aether} becomes $u_\mu=\delta_\mu^T/\sqrt{g^{TT}}$. Within this approach, the two actions given in Eqs. \eqref{AccionHorava} and \eqref{eq:action_ae} can be related to each other if the following relations among the parameters hold \cite{Jacobson:2007veq}:
\begin{equation}
    \frac{G_H}{G_{\ae}} = \xi = \frac{1}{1 - c_{13}}, \qquad
\frac{\lambda}{\xi} = 1 + c_2, \qquad
\frac{\eta}{\xi} = c_{14}
\end{equation}
where the combination $c_{ij}$ is defined as $c_{ij} = c_i + c_j$. 

\section{The equations of motion in our framework}

We study the following static and spherically symmetric space–time:
\begin{equation} \label{metrica}
    ds^{2} = f(r)\,dt^{2} - \,\frac{dr^{2}}{f(r)} - r^{2}\,d\Omega_{2}
\end{equation}
where $d\Omega_{2}$ corresponds to the transversal section of a two–sphere. Our line element \eqref{metrica} allows us to study black hole geometries in which the Killing horizon is identified by the conditions $g_{tt}=0$ and $g_{rr}^{-1}=0$. For this reason, below we specialize the field equations of Refs.~\cite{Vernieri:2017dvi,Vernieri:2018sxd}, originally derived for the case $g_{tt}\neq -(g_{rr})^{-1}$, to the line element given in Eq.~\eqref{metrica}. Furthermore, we study the following anisotropic energy--momentum tensor $T_{\mu\nu}=(\rho+p_{\theta})\,v_{\mu}v_{\nu}-p_{\theta}\,g_{\mu\nu}
+\left(p_{r}-p_{\theta}\right)\chi_{\mu}\chi_{\nu}$, where the four-velocity of the fluid is $v^{\mu}=v^t=1/\sqrt{f(r)} $ and the unit spacelike radial vector is $\chi^{\mu}=\chi^{r}=\sqrt{f(r)}$. This leads to
\begin{equation} \label{TensorEM}
T^{\mu}{}_{\nu}=\mathrm{diag}\big(\rho(r),- p_{r}(r),- p_{\theta}(r),- p_{\theta}(r)\big).
\end{equation}

The {\ae}ther vector field, which is timelike and normalized to unity by definition, becomes hypersurface-orthogonal under spherical symmetry. Its most general expression can be written as
\begin{equation} \label{AetherGral}
    u^{\alpha} = \left( F(r),\, f(r)\sqrt{F(r)^2-1},\, 0,\, 0 \right),
\end{equation}
where $F(r)$ denotes a generic function. Following Refs.~\cite{Panotopoulos:2020uvq,Vernieri:2017dvi,Vernieri:2018sxd,
Vernieri:2019vlh}, we consider the case of a static {\ae}ther aligned with the four-velocity of the matter fluid, so that an observer comoving with the matter is also comoving with the {\ae}ther.
\begin{equation} \label{Aether}
    u^{\alpha} = \left( \frac{1}{\sqrt{f(r)}},\, 0,\, 0,\, 0 \right).
\end{equation}
As can be seen from Eq.~\eqref{AetherGral}, this choice does not correspond to the most general static and spherically symmetric configuration, since the radial component of the {\ae}ther field can be nonzero. However, in line with the aim of the present work, namely to obtain exact analytical solutions for linear equations of state, this simplification is convenient, since the inclusion of a radial component considerably increases the complexity of the field equations, making the construction of exact analytical solutions within the procedure adopted in this work significantly more difficult.

In the present work, we focus on the geometry and thermodynamics of the event horizon. In this regard, we note that, in the case of a black hole geometry with $f(r_h)=0$, the temporal component of the {\ae}ther field diverges at $r=r_h$. For the static {\ae}ther configuration given by Eq.~\eqref{Aether}, the {\ae}ther Lagrangian defined in Eq.~\eqref{eq:lagrangian_ae} reduces to
\begin{equation}
    \mathcal{L}_{\ae}=c_{14}\,a_\mu a^\mu \sim a_\mu a^\mu ,
\end{equation}
where the \ae ther four-acceleration (also known as the {\ae}ther acceleration invariant) is defined as $a^\mu=u^\nu\nabla_\nu u^\mu$. A straightforward calculation yields $a^\mu=\left(0,\frac{f'(r)}{2},0,0\right)$, where a prime denotes differentiation with respect to the radial coordinate. Consequently, the corresponding {\ae}ther acceleration invariant is given by
\begin{equation} \label{InvarianteAether}
    a_\mu a^\mu
    =
    -\frac{1}{4f(r)}
    \left(\frac{df(r)}{dr}\right)^2.
\end{equation} The presence of the metric function $f(r)$ in the denominator of the above expression suggests that the {\ae}ther acceleration invariant may diverge at the event horizon. We first consider the extremal case, for which the event horizon is $m$-fold degenerate, with $m \ge 2$ being a positive integer. In the vicinity of the horizon, the metric function can be written in the generic form $f(r)\sim A(r-r_h)^m$, where $A$ is a nonzero constant. Consequently $a_\mu a^\mu
\sim
-(A/4)\,m^2\,(r-r_h)^{m-2}$.
Therefore, one finds that
\begin{equation} \label{CasosAether}
a_\mu a^\mu
\rightarrow
\left\{
\begin{array}{ll}
\infty, & m=1,\\[2mm]
-A, & m=2,\\[2mm]
0, & m>2.
\end{array}
\right.
\end{equation} Thus, it is straightforward to verify from Eq.~\eqref{CasosAether} that the {\ae}ther acceleration invariant diverges at the event horizon for the nondegenerate case, whereas it remains finite for the degenerate one. As will be seen below, our second case study always corresponds to an extremal black hole, whereas the first and third cases may also become extremal depending on the choice of parameters.  In order to investigate the role of the {\ae}ther acceleration invariant, we write the field equations in terms of this invariant, obtaining

\begin{equation} \label{ComponenteT}
\frac{\eta}{\xi} \left(-\frac{1}{2}\frac{d^2f(r)}{dr^2}-\frac{1}{r}\frac{df(r)}{dr}\right)-\frac{1}{r}\frac{df(r)}{dr}
-\frac{f(r)}{r^2}+\frac{1}{r^2} -\frac{\eta}{2\xi}\,a_\mu a^\mu=8\pi G_{\ae}\rho(r).
\end{equation}

\begin{equation} \label{ComponenteR}
\frac{1}{r}\frac{df(r)}{dr}
+\frac{f(r)}{r^2}
-\frac{1}{r^2} -\frac{\eta}{2\xi}\,a_\mu a^\mu
=8\pi G_{\ae}p_r(r).
\end{equation}

\begin{equation} \label{ComponenteTheta}
\frac{1}{2}\frac{d^2f(r)}{dr^2}
+\frac{1}{r}\frac{df(r)}{dr} + \frac{\eta}{2\xi}\,a_\mu a^\mu
=8\pi G_{\ae}p_\theta(r).
\end{equation}

As can be seen from Eqs.~\eqref{ComponenteT}--\eqref{ComponenteTheta}, once the field equations are expressed in terms of the {\ae}ther acceleration invariant, the latter appears explicitly as a contribution to the gravitational sector. Consequently, the energy--momentum components can be decomposed as
\begin{align}
8\pi G_{\ae}\rho(r)
&=
8\pi G_{\ae}\rho_0(r)
-\frac{\eta}{2\xi}\,a_\mu a^\mu,
\\
8\pi G_{\ae}p_r(r)
&=
8\pi G_{\ae}p_{r0}(r)
-\frac{\eta}{2\xi}\,a_\mu a^\mu,
\\
8\pi G_{\ae}p_\theta(r)
&=
8\pi G_{\ae}p_{\theta0}(r)
+\frac{\eta}{2\xi}\,a_\mu a^\mu,
\end{align}
First, we note that, in the degenerate case, the matter sources remain regular at the event horizon. Second, for the nondegenerate case, we find that the {\ae}ther acceleration invariant is directly associated with the divergent contribution to the matter sources. Accordingly, the functions $\rho_0(r)$, $p_{r0}(r)$, and $p_{\theta0}(r)$ denote the contributions that remain finite at the horizon. The latter is straightforward to verify since, as will be seen below for the three case studies considered in this work, the metric function $f(r)$, together with its first and second derivatives with respect to the radial coordinate, remains finite there. Therefore, the near-horizon behavior of the matter variables is directly determined by that of the {\ae}ther acceleration invariant. As discussed above, this invariant diverges for nondegenerate event horizons. Consequently, the matter variables inherit this divergence through the field equations, with both sides of the field equations exhibiting the same leading-order divergent behavior near the event horizon. Thus, the divergence of the static {\ae}ther ansatz at a nondegenerate event horizon is not merely a coordinate artifact. Rather, within the strictly static {\ae}ther configuration adopted in this work, it may be interpreted as a potential pathology of the {\ae}ther sector. On the other hand, the spacetime geometry itself remains regular, since it is straightforward to verify that the Ricci and Kretschmann curvature invariants remain finite at the event horizon. This suggests that the potential pathology is associated with the static {\ae}ther configuration rather than with the spacetime geometry. Whether a more general {\ae}ther configuration can remove this potential pathology while preserving the same geometry remains an open question and lies beyond the scope of the present work. It is also worth investigating whether, and to what extent, this potential pathology affects the interpretation of Killing-horizon thermodynamics. Although the Hawking temperature and the Wald entropy remain finite and can be computed within the standard formalism, the physical implications of the divergent {\ae}ther invariant and effective matter variables for the thermodynamics of these solutions remain to be clarified. Interestingly, this potential pathology is absent in the extremal solutions discussed in Cases I, II, and III, for which the event horizon is degenerate. In these configurations, the {\ae}ther acceleration invariant, the effective matter sources, and the geometric invariants remain finite at the horizon, while the Einstein--{\ae}ther field equations remain free of divergences. A complete understanding of the extremal horizon structure in these solutions deserves further investigation.

It is worth mentioning that, in the nondegenerate case, the {\ae}ther field becomes imaginary in the interval $r_i<r<r_h$,  where $r_i$ denotes the inner horizon, whereas in the degenerate case this occurs for $r<r_{\rm ext}=r_h$. This behavior would seem to indicate that it could not be possible to extend the preferred foliation associated with the {\ae}ther field through these regions, although a deeper analysis is required, which lies beyond the scope of the present work. In this context, as mentioned previously, the present work focuses on the physics at the event horizon and the exterior geometry, where the {\ae}ther field remains real. On the other hand, In our framework, the conservation equation of the energy-momentum tensor takes the form:
\begin{equation}
    p_{r}'(r) + \frac{[\rho(r) + p_{r}(r)]\, f'(r)}{2\, f(r)} = \frac{2}{r}\,[p_{\theta}(r) - p_{r}(r)]
\end{equation}
There are four equations of motion, however, only three of the above equations are actually independent. In the set of equations above, the effective contributions to the energy density and pressures arising from the {\ae}ther are determined by the parameter $\eta/\xi$. General Relativity (GR) is naturally recovered when $\eta = 0$. As previously noted, the analysis focuses on the case of a static {\ae}ther. According to the authors of \cite{Vernieri:2017dvi}, if this condition is not satisfied, two additional equations must be taken into account. It is also worth emphasizing that, despite the differences in the general field equations, in this specific scenario (spherical symmetry with a static {\ae}ther), the resulting equations exactly coincide with those obtained in the Einstein--{\ae}ther theory \cite{Blas:2010hb}.

\section{Our Black hole solutions with a Linear Equation of State}

In the equations of motion described above, we can notice that, under a metric tensor of the form $g_{tt}=-g_{rr}^{-1}$, Eq. \eqref{metrica} no longer satisfies the condition $T^{t}_{t}=T^{r}_{r}\Rightarrow \rho=-p_r$. This latter condition is highly typical of black hole solutions sourced by matter fields, both in General Relativity and in some of its extensions, and can therefore be regarded in those cases as an equation of state implicitly encoded in the equations of motion. In this way, the fact that the mentioned equation of state is modified by the Hořava {\ae}ther terms motivates us to test alternative equations of state for black hole solutions and to analyze how these terms influence the physical properties of black holes. This latter effect will also be tested at short scales in our last case study. In accordance with the strategy described in the Introduction, we seek linear EoS that lead directly to exact analytical black hole solutions. In the following, we analyze the physically relevant scenario associated with each equation of state together with the geometrical and physical properties of the resulting spacetimes. In this context, we first note that the most general linear equation of state, $N_1\rho(r)+N_2p_r(r)+Np_\theta(r)=0$, with $N_1\neq0$, $N_1\neq1$, $N_2\neq0$, $N_2\neq1$, and $N\neq0$, does not seem to admit exact analytical solutions within the procedure adopted in this work. Although we do not exclude the existence of more sophisticated analytical or numerical solutions for this more general case, here we focus on the three particular cases discussed in the following subsections, which do lead directly to exact analytical black hole solutions. On the other hand, as discussed in the Introduction, we also investigate some thermodynamic properties of the resulting black hole solutions. In particular, we analyze how the different linear EoS influence the thermal behavior of these solutions and the possible existence of extremal black holes. The associated Hawking temperature $T_H$ is determined by the surface gravity $\kappa$, defined through $\kappa^2=-\frac{1}{2}\nabla^\mu k^\nu \nabla_\mu k_\nu \big|_{r=r_h}$, where $k^\mu$ is the timelike Killing vector. Thus, $T_H=\kappa/(2\pi)$. For the line element considered in this work, the temperature reduces to $T_H=(4\pi)^{-1}df/dr\big|_{r=r_h}\,$.

\subsection{Case I: Analogue charged black hole}

It is well known that the Reissner--Nordström (RN) spacetime constitutes a black hole geometry whose matter sources satisfy the relation $\rho(r)=-p_r(r)=p_\theta(r)$, which directly leads to the linear EoS
\begin{equation} \label{EoSRN}
p_r(r)+p_\theta(r)=0.
\end{equation}
It is worth mentioning that, in this work, we do not derive this EoS from a specific Lagrangian. Rather, we adopt it as a linear EoS motivated by the fact that the pressure components associated with the RN black hole solution in General Relativity naturally satisfy this relation. In particular, by substituting the expressions for $p_r$ and $p_\theta$, given by Eqs.~\eqref{ComponenteR} and \eqref{ComponenteTheta}, respectively, into Eq.~\eqref{EoSRN}, we obtain the solution
\begin{equation}
\label{SolucionAnalogaRN}
f(r)=1-\frac{C_1}{r}+\frac{C_2}{r^2}.
\end{equation}

The above solution formally coincides with the RN metric upon identifying $C_1=2M$ and $C_2=q^2$, where $M$ and $q$ represent the mass and electric charge, respectively. However, the matter sources supporting this geometry do not satisfy the characteristic General Relativity relation $\rho=-p_r$, which appears both in the Maxwell electromagnetic stress--energy tensor and in several extensions based on nonlinear electrodynamics. Consequently, the matter sources generated by HG--EA gravity can be interpreted as an exotic anisotropic matter distribution induced by the modified gravitational dynamics, which gives rise to a geometric term analogous to an electric potential in the solution.  In agreement with the general discussion presented in Section~III, the {\ae}ther acceleration invariant is given by
$a_\mu a^\mu=-\frac{(Mr-Q^2)^2}{r^6\left(r^2-2Mr+Q^2\right)}$.
Consequently, the matter variables take the form
\begin{align}
8\pi G_{\ae}\rho(r)
&=
\left(1-\frac{\eta}{\xi}\right)\frac{Q^2}{r^4}
-\frac{\eta}{2\xi}\,a_\mu a^\mu,
\\
8\pi G_{\ae}p_r(r)
&=
-\frac{Q^2}{r^4}
-\frac{\eta}{2\xi}\,a_\mu a^\mu,
\\
8\pi G_{\ae}p_\theta(r)
&=
\frac{Q^2}{r^4}
+\frac{\eta}{2\xi}\,a_\mu a^\mu.
\end{align}
First, we note that, for the doubly degenerate case with $M=Q$ and the extremal event horizon $r_h=M$, the {\ae}ther acceleration invariant at the horizon is given by $a_\mu a^\mu=-M^{-4}$, which is finite. Consequently, it is straightforward to verify that, in this case, the matter sources also remain finite at the horizon.

For the nondegenerate case, the divergent behavior at the event horizon is entirely encoded in the contribution proportional to the divergence of the {\ae}ther acceleration invariant, while the remaining finite contributions to the pressure components arise from the standard electromagnetic sector of the Reissner--Nordstr\"om solution. In contrast, the energy density receives an additional finite correction proportional to $\eta/\xi$. Since the geometry obtained in this subsection resembles the RN solution, whose thermodynamics is well known, we omit its thermodynamic analysis here and instead focus on the thermodynamic properties of the following case studies.

\subsection{Case II: Extremal black hole}

In accordance with the strategy described in the Introduction and at the beginning of this section, in the present subsection we consider a equation of state of the form
\begin{equation} \label{EcuacionDeEstadoExtremo}
    \rho(r)= - p_{r}(r) -N \cdot p_{\theta}(r)
\end{equation}
where $N$ is a real parameter. Once the solution is obtained, we discuss the physically meaningful scenario associated with this EoS, as well as the geometrical and physical properties of the resulting spacetime. As will be seen below, $N$ is directly related to the modification introduced by the HG--EA sector, more specifically $N\sim\eta$. In the limit $N\to0$, the General Relativity relation $\rho=-p_r$ is recovered. The physical implications of this parameter will be discussed through the geometrical and physical properties of the resulting spacetime.

By substituting the expressions for $\rho$, $p_r$, and $p_\theta$ given by equations \eqref{ComponenteT}, \eqref{ComponenteR}, and \eqref{ComponenteTheta}, respectively, into equation \eqref{EcuacionDeEstadoExtremo}, we obtain a solution of the form:
\begin{equation} \label{SolucionConN}
f(r)=\left(C_1+\frac{C_2}{r}\right)^{\dfrac{4\eta-4N\xi}{(N+2)\eta-4N\xi}
}.
\end{equation}
where $(N+2)\eta-4N\xi \neq 0$. We choose $C_1=1$ and $C_2=-2M$ in order to compare the solution with the Schwarzschild geometry. For $N=2$, the exponent becomes unity and the geometry resembles the usual Schwarzschild solution for arbitrary values of $\eta$ and $\xi$. For $N\neq2$ and $N\neq0$, the metric function does not reduce to the Schwarzschild form for generic values of $\eta$ and $\xi$, making explicit the influence of the HG--EA sector on this solution. Although $N=0$ is not included in the class of exact solutions considered here, it represents the General Relativity limit through the relation $N\sim\eta$. In the limit ${\xi,\eta}\rightarrow{1,0}$, the exponent tends to unity and the geometry continuously approaches the Schwarzschild solution. In the same limit, the corresponding matter sector reduces to the vacuum case, $\rho=p_r=p_\theta=0$. In the following, we consider only the cases with $N\neq2$ and $N\neq0$, together with the assumptions described below. On the other hand, in order to ensure a change of signature, we identify two cases:
 \begin{itemize}[align=left]
  \item The exponent of equation \eqref{SolucionConN} corresponds to a fraction whose denominator is an integer, positive and odd value: 
     \begin{equation}
      \dfrac{1}{\bar{n}_{\text{odd}}}=  \dfrac{4\eta-4N\xi}{(N+2)\eta-4N\xi}
    \end{equation}
First, in accordance with the discussion presented above, we note that $N\sim\eta$, more specifically, $N=2\eta\,(2-\bar{n}_{\text{odd}}^{-1})/(\bar{n}_{\text{odd}}^{-1}\,\eta+4\xi-4\xi\,\bar{n}_{\text{odd}}^{-1})$. On the other hand, in accordance with the discussion presented at the beginning of this section, the Hawking temperature, $T_H=(4\pi)^{-1}df/dr\big|_{r=r_h}$, is given by
    \begin{equation}
        T_H \sim \frac{(4\pi)^{-1} \,2M\,\bar{n}_{\text{odd}}^{-1}\,r_h^{-2}}{\left( 1- \frac{2M}{r_h} \right )^{1-\dfrac{1}{\bar{n}_{\text{odd}}}}}
    \end{equation}
    where we observe that the temperature is not well defined at the event horizon $r_h = 2M$, since it diverges for $\bar{n}_{\text{odd}}>1 \in [3,5,7\ldots]$. Therefore, we will not analyze this particular case in this work.
       \item The exponent of equation \eqref{SolucionConN} takes an integer, positive and odd value: 
         \begin{equation}
      n_{\text{odd}}=   \dfrac{4\eta-4N\xi}{(N+2)\eta-4N\xi}
    \end{equation}
 where $n_{\text{odd}}\in [1, 3,5,7\ldots]$.   or, equivalently,
 \begin{equation} \label{ValorDeN}
     N= \frac{2 \eta \cdot (2-n_{\text{odd}})}{n_{\text{odd}} \cdot \eta + 4 (1-n_{\text{odd}}) \xi}
 \end{equation}
 where we note that $N \neq 0$ since, as mentioned above, $n_{\text{odd}} \neq 2$. As mentioned at the beginning of this subsection, the parameter $N$ is related to the Hořava terms. In the remainder of this subsection, we will continue analyzing this case.
\end{itemize}

\paragraph*{\bf Integer exponent-- Extremal black hole and its thermodynamics:} The solution is given by

\begin{equation}
    f(r) = \left ( 1 - \frac{2M}{r}  \right)^{n_{\text{odd}}}
\end{equation}
where, in connection with the discussion above, the Schwarzschild solution is recovered for $N = 2 \Rightarrow n_{\text{odd}} = 1$. We point out the following at the location of the event horizon for $n_{\text{odd}} > 1 \in [3,5,7,\ldots]$ :
\begin{align}
f(r_h)=0=&\left(1 - \frac{2M}{r_h}\right)^{n_{\text{odd}}} \nonumber \\
=&\underbrace{\left(1 - \frac{2M}{r_h}\right) \cdot \left(1 - \frac{2M}{r_h}\right) \cdot \ldots \cdot \left(1 - \frac{2M}{r_h}\right)}_{n_{\text{odd}}  \text{  times}}  
\end{align}

Thus we note that, in this case, the value of the event horizon is $n_{\text{odd}}$-fold degenerate. In agreement with the discussion presented in Section~III, the {\ae}ther acceleration invariant takes the form $a_\mu a^\mu=-\frac{M^2n_{\mathrm{odd}}^2}{r^4}\left(1-\frac{r_h}{r}\right)^{\,n_{\mathrm{odd}}-2}$. For the values considered here, namely $n_{\mathrm{odd}}\in\{3,5,7,\ldots\}$, this invariant vanishes at the degenerate event horizon. Consequently, the matter variables evaluated at the horizon are
\begin{align}
8\pi G_{\ae}\rho(r_h)
&=
\frac{1}{r_h^2},
\\
8\pi G_{\ae}p_r(r_h)
&=
-\frac{1}{r_h^2},
\\
8\pi G_{\ae}p_\theta(r_h)
&=
0.
\end{align}
Therefore, both the {\ae}ther acceleration invariant and the matter variables remain finite at the degenerate event horizon. In particular, the matter sector satisfies $\rho(r_h)=-p_r(r_h)$, while the tangential pressure vanishes at the event horizon.

The temperature is given by
 \begin{equation}
        T_H \sim \frac{2M\,n_{\text{odd}}}{4 \pi r_h^2} \left ( 1- \frac{2M}{r_h}  \right )^{n_{\text{odd}}-1}
    \end{equation}
    
It is straightforward to note that the degeneracy of the event horizon leads to the temperature vanishing for $n_{\text{odd}} > 1$, with $n_{\text{odd}} \in {3,5,7,\ldots}$.  Therefore, in this latter case we are dealing with an extremal black hole. Thus, such degeneracy in the root of the function $f(r)$ implies that $f'(r_{h}) = 0$, and therefore its temperature vanishes, $T_H = 0$. That is, extremal black holes do not emit Hawking radiation. Nevertheless, they do possess entropy, since it depends only on the number of quantum states of the system.

The form of the spacetime, together with the action principle for gravity, allows one to define the thermodynamics of these solutions. This, in turn, makes it possible to compute the entropy as part of the Noether charge on the horizon, following Wald’s original approach \cite{Wald:1993nt}. In this case, the entropy is given by
    \begin{equation} \label{EntropiaSinLimite}
        S = \frac{ Q(\xi)}{T_H}\Big|_{r=r_h}
    \end{equation}
where $\xi = \xi^{\mu}\partial_{\mu}$ is the vector field that generates the diffeomorphism. In our case, $\xi = \xi^{t}\partial_{t} = \partial_{t}$ is a timelike vector, with $\xi^{t} = (1,0,0,0)$ also being timelike. 
Since both the Noether charge $Q(\partial_t)$ and the temperature are evaluated at the horizon, and since the temperature vanishes in our case, the entropy is then defined as:

 \begin{equation} \label{Entropia}
       S = \displaystyle \lim_{r \to r_h}   \frac{ Q(\partial_t)}{T_H}
    \end{equation}

In order to compute the Noether charge, we use the Komar formula \cite{Komar:1958wp}. As shown in Ref. \cite{Aros:1999kt}, this expression can also be associated with the Noether conserved charge, including boundary terms in the action, which, in the absence of a cosmological constant,  leads to the conserved charge being twice the value obtained from the Komar formula.
\begin{equation}
    \lim_{r \to r_h} Q(\partial_t)
= \lim_{r \to r_h} \frac{1}{16\pi} \frac{df}{dr}\!\cdot  r^{2} \int d\Omega_{2}= 
T_H \cdot \frac{r_h^2 \cdot 4 \pi}{4} =  T_H \cdot \frac{\mbox{area}}{4}
\end{equation}
Substituting into Equation \eqref{Entropia}
\begin{equation} \label{Area}
    S= \frac{\mbox{area}}{4}
\end{equation}

We denote by $\ell$ the unit of length. Accordingly, the geometrized mass has dimensions $[M]=\ell$, so that the ratio $M/r$ is dimensionless. Moreover, consistency with the usual first law of thermodynamics, $dM=T_H\,dS$, together with $[T_H]=\ell^{-1}$, requires the entropy to have dimensions $[S]=\ell^2$. This is consistent with the entropy expression in Eq. \eqref{Area} when geometrized units are adopted \cite{CamaradS:2013pcy}. Thus, following Wald’s procedure, we find that the entropy obeys the area law. This is a non-trivial result, since, when other methodologies are employed, the entropy of black holes in the presence of matter usually does not follow the area law \cite{Ma:2014qma}. 

\subsection{Case III: Equation of state analogous to an ultra--relativistic stiff fluid}
In accordance with the strategy described in the Introduction and at the beginning of this section, in the present subsection we consider an equation of state of the form
\begin{equation} \label{EcuacionDeEstadoUltraRelativista}
      \rho(r)=p_{r}(r)
\end{equation}

This equation of state corresponds to an ultrarelativistic stiff fluid. It was first proposed by Zeldovich \cite{Zeldovich:1983cr} in a cosmological setting. As emphasized in Ref. \cite{Ray:2020yyk}, such an equation of state can be interpreted in terms of “soft quanta”, meaning that it models simple quantum excitations that effectively represent an ultrarelativistic stiff fluid without requiring a detailed description of the underlying microphysics at extreme densities. The same reference also notes that the stiff-fluid paradigm has been employed in both astrophysics and cosmology on multiple occasions to characterize high-density matter. This kind of fluid lies at the causal limit, since the speed of sound reaches the speed of light. In the gravastar framework \cite{Mazur:2001fv}, this equation of state is used to model a layer of stiff matter, commonly referred to as the shell, which is thin yet has a finite thickness.

By substituting the expressions for $\rho$ and $p_r$ given by equations \eqref{ComponenteT} and \eqref{ComponenteR}, respectively, into equation \eqref{EcuacionDeEstadoUltraRelativista}, we obtain a solution of the form:
\begin{equation}
    f(r)=1-\frac{C_1}{r}+ \frac{C_2}{r^n}
\end{equation}
where $n=4 \xi/\eta=4/c_{14}$. For $C_1=2M$ and $C_2>0$, where $M$ represents the mass, the metric function represents the Schwarzschild metric plus a repulsive potential $C_{2}/r^{4\xi/\eta}$. We denote by $\ell$ the unit of length. Accordingly, the geometrized mass has dimensions $[M]=\ell$, while $[C_2]=\ell^{\,n}$. We note that this repulsive potential ensures that the metric remains asymptotically flat. In the special case where $n=2 \Rightarrow \xi/\eta = 1/2$ and $C_{2} = q^{2}$, the metric also resembles the Reissner–Nordström form, and therefore the physical arguments discussed after Eq. \eqref{SolucionAnalogaRN} could also apply to this special case. It is also worth mentioning that for $n=3 \Rightarrow\xi/\eta = 3/4$ the correction to the Newtonian potential resembles that obtained from the GUP parameter arising from quantum corrections \cite{Scardigli:2016pjs}. For $n=4 \Rightarrow\xi/\eta = 1$ it resembles the quantum correction in a (pseudo) static, spherically symmetric semiclassical Oppenheimer–Snyder model \cite{Lewandowski:2022zce}.

\begin{figure}[ht]
\centering

\begin{subfigure}{\textwidth}
    \centering
    \includegraphics[width=3.6in]{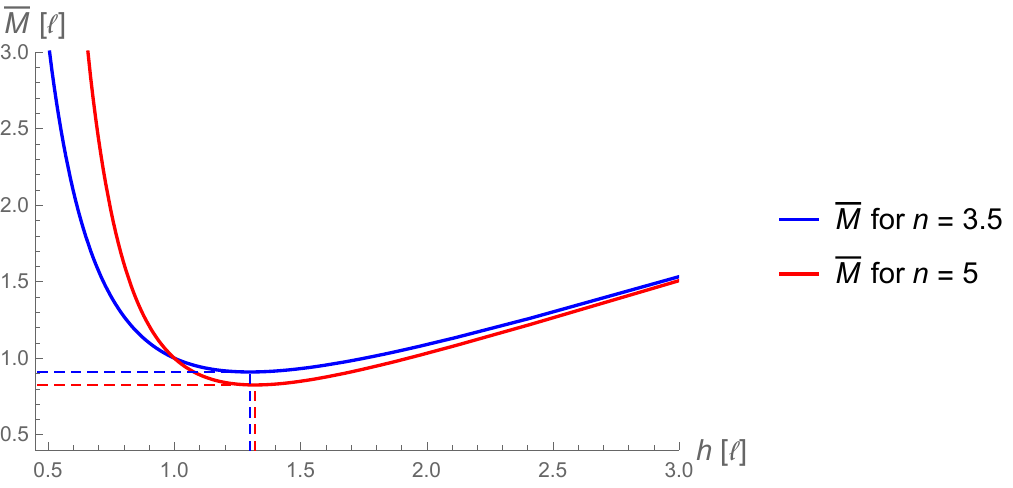}
    \caption{Mass parameter with units $[\bar{M}]=\ell$ versus $h$, with units $[h]=\ell$, obtained from the condition $f(h,\bar{M})=0$, for $C_2=1$ and $n=4\xi/\eta=3.5,\,5$.}
    \label{fig:masaJulio20}
\end{subfigure}

\begin{subfigure}{\textwidth}
    \centering
    \includegraphics[width=4.2in]{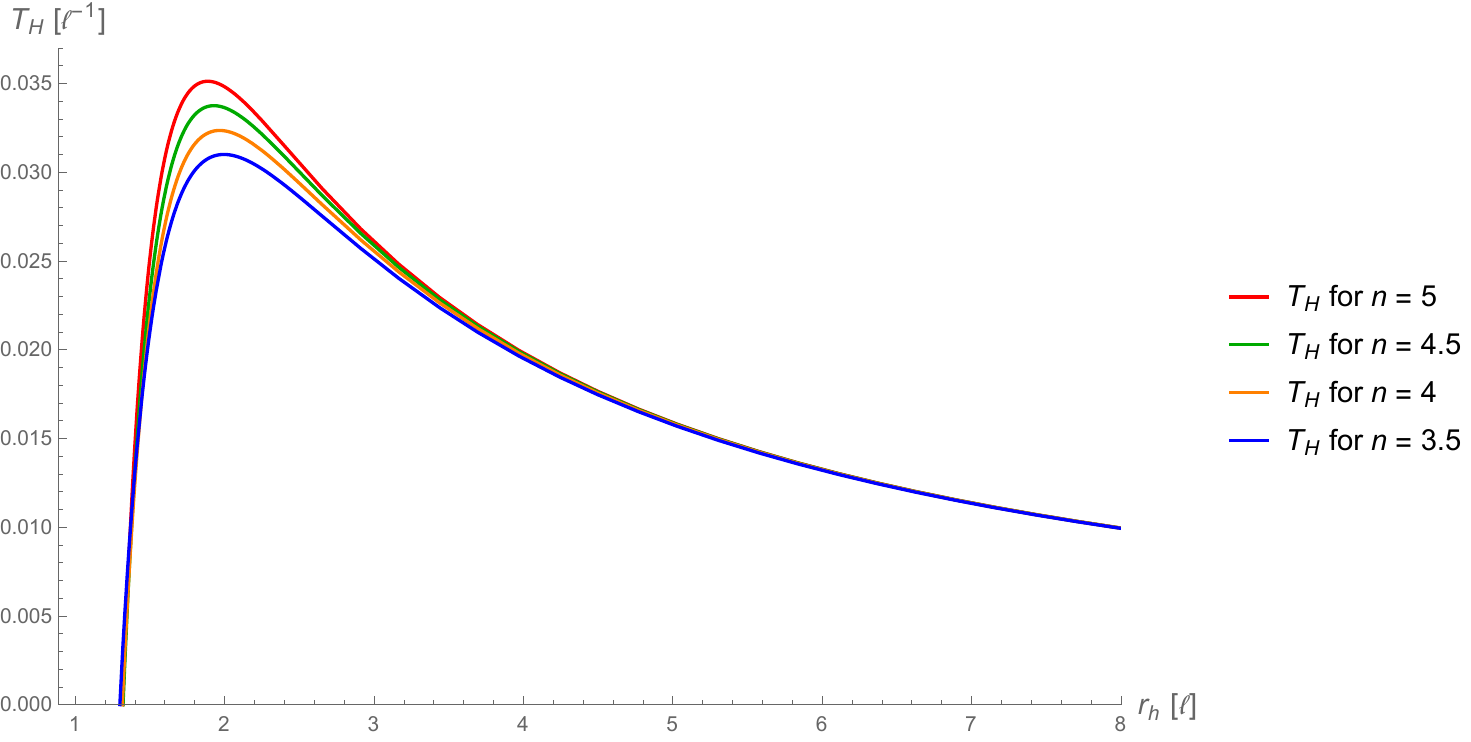}
    \caption{Temperature with units $[T_H]=\ell^{-1}$ versus the event horizon radius with units $[r_h]=\ell$ for $C_2=1$ and $n=4\xi/\eta=3.5,\,4,\,4.5,\,5$.}
    \label{fig:tempJulio20}
\end{subfigure}

\begin{subfigure}{\textwidth}
    \centering
    \includegraphics[width=4.2in]{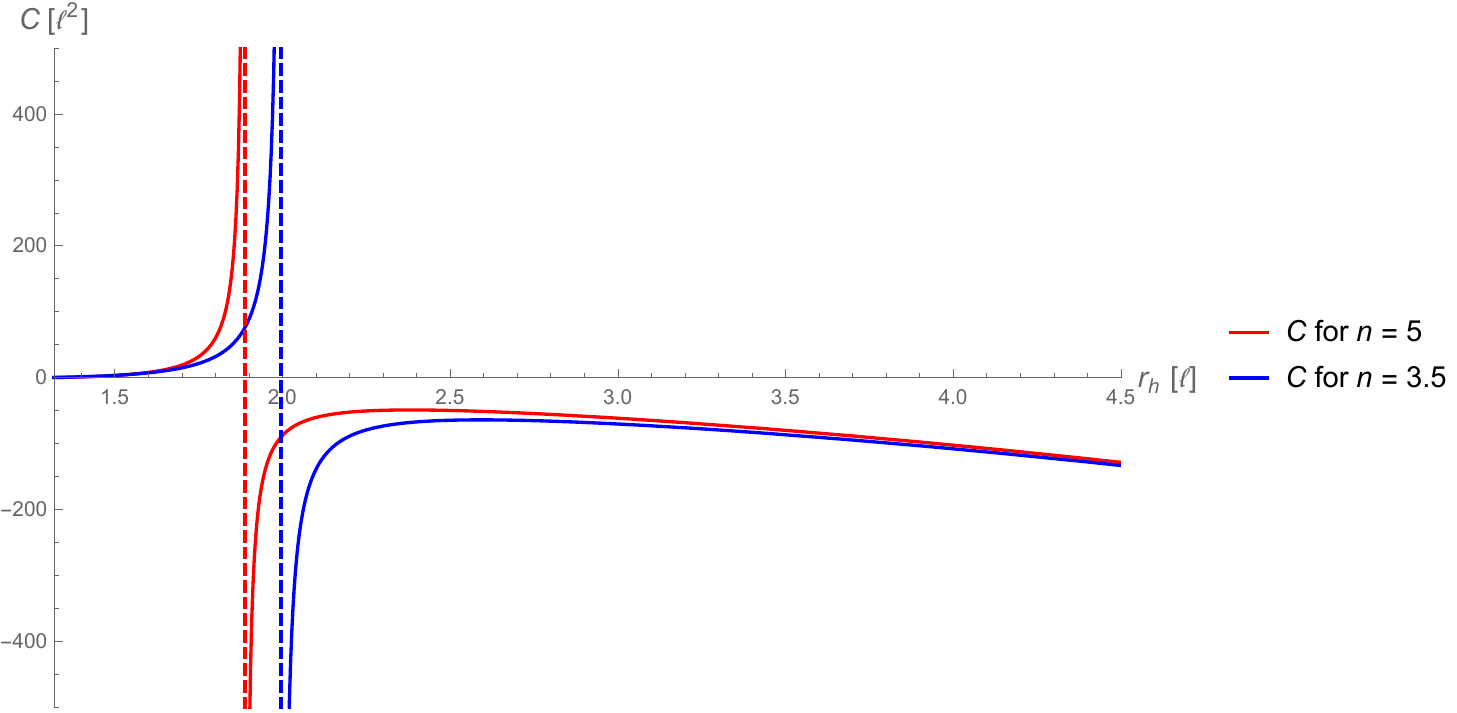}
    \caption{Heat capacity with units $[C]=\ell^{2}$ versus the event horizon radius with units $[r_h]=\ell$ for $C_2=1$ and $n=4\xi/\eta=3.5,\,5$.}
    \label{fig:capJulio20}
\end{subfigure}

\caption{Thermodynamic behavior.}
\label{FigTemperatura}

\end{figure}

In order to study the horizon structure, we define the mass parameter $\bar{M}$ as the value of the parameter $M$ satisfying the condition $f(r=h,M=\bar{M})=0$, yielding
\begin{equation}
    \bar{M}= \frac{h}{2} + \frac{C_2}{2h^{n-1}}.
\end{equation}where, according to the convention introduced above, the dimensions are $[\bar{M}]=\ell$, $[h]=\ell$, and $[C_2]=\ell^{\,n}$. A generic behavior of this function is shown in panel (a) of Fig.~\ref{FigTemperatura}. We observe that the curve possesses a minimum corresponding to the extremal point $(h_{\mathrm{ext}},M_{\mathrm{ext}})$. The points located to the left of $h_{\mathrm{ext}}$, namely $h=r_-$, are associated with a potential inner horizon, whereas the points located to the right of the minimum, $h=r_h>h_{\mathrm{ext}}$, correspond to the event horizon. For $M>M_{\mathrm{ext}}$, the curve therefore presents two branches: $h=r_-<h_{\mathrm{ext}}$ and $h=r_h>h_{\mathrm{ext}}$, corresponding to the event horizon. Since the present analysis is restricted to the event-horizon branch, we investigate the Hawking temperature and the heat capacity only for the event horizon. The extremal point can be obtained analytically by imposing the condition $\frac{d\bar{M}}{dh}=0$, which yields
\begin{equation} \label{EcuacionExtremo}
    (h_{\text{ext}}, M_{\text{ext}})= \left ( \left (C_2(n-1) \right )^{1/n}, \frac{C_2 \cdot n}{2} \left (C_2(n-1) \right )^{1/n-1} \right )
\end{equation} Since $h_{\mathrm{ext}}$ represents the extremal radius and $M_{\mathrm{ext}}$ the mass parameter of the extremal black hole, we will consider physically admissible only those cases in which both quantities are real and positive in Eq.~\eqref{EcuacionExtremo}. For this reason, in our analysis we consider values such that $C_2>0$ and $n>1$. It is worth mentioning that we do not consider cases with $C_2>0$ and $n=1/2,1/4,1/6,\ldots$, since these can lead to negative values of the extremal mass. For example, $C_2=1$ and $n=1/2$ lead to $M_{\mathrm{ext}}=-1/8<0$.  As mentioned, the values $(h_{\text{ext}}, M_{\text{ext}})$ describe the extremal black hole, where, as we will discuss below, the temperature vanishes and a black-hole remnant is formed.

In this subsection, we consider an ultra-relativistic scenario that, at present, appears to be viable mainly from a theoretical point of view. For this reason, a more complete discussion of the allowed parameter values requires a deeper analysis, which is beyond the scope of the present work. Nevertheless, it is useful to comment on some known constraints on the parameter $c_{14}$, which in our case is related to $n$ through $n=4/c_{14}$. Refs.~\cite{Arata:2026iec,Berglund:2012bu} indicate that, in order to avoid pathologies such as negative-energy modes and to ensure an attractive Newtonian potential, the condition $0<c_{14}<2$ must be satisfied, implying $n>2$. Therefore, the values $n=3$ and $n=4$, which satisfy the above condition, should be regarded only as illustrative examples within the theoretical ultra-relativistic scenario considered here, rather than, as discussed below, as values selected to satisfy current observational constraints. On the other hand, studies following GW170817, see for example Ref.~\cite{Oost:2018tcv} and references therein, have proposed much more stringent observational constraints, such as $0<c_{14}\leq2\times10^{-7}$ in certain regions of the parameter space. In our case, this would imply very large values of $n$, namely $n>2\times10^{7}$. As mentioned above, a detailed analysis of these constraints within the ultra-relativistic scenario considered here requires a more in-depth investigation and could be addressed in future work.

In agreement with the discussion presented in Section~III, the matter variables are given by
\begin{align}
8\pi G_{\ae}\rho(r)
&=
\frac{C_2(n-1)}{r^{n+2}}
\left(
1-\frac{\eta n}{2\xi}
\right)
-\frac{\eta}{2\xi}\,a_\mu a^\mu,
\\
8\pi G_{\ae}p_r(r)
&=
-\frac{C_2(n-1)}{r^{n+2}}
-\frac{\eta}{2\xi}\,a_\mu a^\mu,
\\
8\pi G_{\ae}p_\theta(r)
&=
\frac{C_2n(n-1)}{2r^{n+2}}
+\frac{\eta}{2\xi}\,a_\mu a^\mu.
\end{align}
In the extremal case, the event horizon $r_h=h_{\text{ext}}$ and the mass parameter $M=M_{\text{ext}}$ are given by Eq.~\eqref{EcuacionExtremo}. Defining $x=r_h/r$, the metric function and its radial derivative are given by $f(r)=(x^n-nx+n-1)/(n-1)$, and $f'(r)=\left[nr_h(1-x^{\,n-1})\right]/\left[(n-1)r^2\right]$. Since the event horizon corresponds to $x=1$, we expand both expressions around this point. Using
$x^n=1+n(x-1)+n(n-1)(x-1)^2/2+\mathcal{O}\!\left((x-1)^3\right)$, it follows that $x^n-nx+n-1=n(n-1)(x-1)^2/2+\mathcal{O}\!\left((x-1)^3\right)$,
whereas $1-x^{\,n-1}=-(n-1)(x-1)+\mathcal{O}\!\left((x-1)^2\right)$. Substituting these expansions into Eq.~\eqref{InvarianteAether} and taking the limit $x\rightarrow1$, the {\ae}ther acceleration invariant at the event horizon becomes $a_\mu a^\mu(r_h)=-n/(2r_h^2)=-n/\left(2[C_2(n-1)]^{2/n}\right)$. Therefore, the {\ae}ther acceleration invariant remains finite at the degenerate event horizon for arbitrary real $n>1$. The effective matter sources given in Eqs.~(45)--(47) likewise remain finite and free of divergences at the horizon.

From Eq.~\eqref{InvarianteAether}, we also note that the {\ae}ther acceleration invariant is given by $a_\mu a^\mu=-\frac{\left[A(r)\right]^2}{4f(r)}$, where $A(r)=2M/r^2-nC_2/r^{n+1}$. Thus, for a nondegenerate event horizon, this invariant diverges at $r=r_h$. Consequently, the matter variables $\rho$, $p_r$, and $p_\theta$ given in Eqs.~(45)--(47) inherit the same divergence at the event horizon. Therefore, the divergent behavior is entirely encoded in the term proportional to the {\ae}ther acceleration invariant.

\paragraph*{ \bf A brief discussion of the thermodynamics of this case:} First, we note that, in the extremal case, the Hawking temperature vanishes due to the degeneracy of the event horizon. Furthermore, according to the definition introduced in Eq.~\eqref{EntropiaSinLimite}, it is straightforward to verify that the entropy satisfies the area law given by Eq.~\eqref{Area} for both the extremal and nonextremal cases. In what follows, we focus on the thermodynamic properties of the non--extremal configuration. As mentioned above, the potential pathology in the non--extremal case is associated with the static {\ae}ther configuration rather than with the horizon geometry itself. Thus, the geometric thermodynamic quantities remain well defined despite the divergence of the static {\ae}ther acceleration invariant. Along these lines, investigating a possible connection between this potential pathology and the thermodynamic properties requires a more in-depth analysis, which lies beyond the scope of the present work. It is worth noting that, in this case, it is not necessary to evaluate limit \eqref{Entropia}, since the temperature does not vanish for all values of $r_h$. The temperature is given by:
\begin{equation}
    T_H= \frac{1}{4\pi} \frac{df}{dr} \Big |_{r=r_h}= \frac{1}{4 \pi r_h} - \frac{C_2(n-1)}{4 \pi r_{h}^{n+1}}
\end{equation}

We note that the first term resembles the Schwarzschild temperature. On the other hand, the second term depends on $n = 4\xi / \eta$, that is, on the parameters of the HG--EA theory, which modify the gravitational field equations. In order to test the influence of these latter terms, we write the derivative of the temperature as follows:

\begin{equation}
   \frac{dT_H}{dr} \Big |_{r=r_h}=  \frac{1}{4 \pi} \left ( -\frac{1}{r_h^2} + \frac{C_2(n-1)(n+1)}{r_h^{2+n}}\right )
\end{equation}

On the one hand, we observe that the first term has a negative slope, resembling the Schwarzschild temperature, which increases without bound as the mass and the horizon radius decrease, i.e. $T^{\mbox{schw}}_H(M,r_h \to 0) \to \infty$. This term becomes dominant for large values of the event horizon. However, we note that the second term has a positive slope. Since, as mentioned above, equation \eqref{EcuacionExtremo}, $C_2 (n - 1) > 0$, this power-law term $\sim 1/r_h^{n+1}$ with $n > 1$ becomes dominant at small scales. This is consistent with the fact that the HG--EA terms are influential at short scales. This effect also has consequences for the evolution of the temperature. In panel (b) of Fig.~\ref{FigTemperatura}, we display the behavior of the temperature for different values of $n=4\xi/\eta$. We observe that the correction to the temperature at small scales, arising from the presence of the HG--EA terms, prevents the temperature from diverging to infinity as in the Schwarzschild case. In this way, the fact that the slope becomes positive at short scales causes the temperature to start decreasing after reaching a maximum, while approaching the value $T_H = 0$. This final value is attained in the previously described extremal case, where the inner and event horizons coincide.

In panel (c) of Fig.~\ref{FigTemperatura}, we display the behavior of the heat capacity, using the definition $C= T_H \, \frac{dS}{dT_H}= T_H \left (\frac{\partial S}{\partial r_h} \right ) \left (\frac{\partial T_H}{\partial r_h} \right )^{-1}$. We note that a phase transition occurs between the unstable branch at large scales ($C < 0$) and the stable branch at short scales ($C > 0$), taking place at the same location where the temperature reaches its peak. We also observe that the phase transition occurs at larger values of the event horizon radius as $n = 4\xi / \eta$ decreases. In this work, we adopt geometrized units, in which the heat capacity has dimensions $[C]=\ell^2$. This is consistent with the standard definitions $C=dM/dT_H$ and $C=T_H\,dS/dT_H$.

From the analysis of the temperature and the heat capacity we can see that, as the event horizon approaches small scales, the black hole becomes stable. This is due to the correction to the Schwarzschild temperature that arises at short distances from the presence of the HG--EA terms. This correction, besides preventing the temperature from diverging to infinity, causes it to decrease and vanish at $T_H = 0$. The latter corresponds to the extremal case. In this situation, a black-hole remnant is formed, which can be interpreted as what remains of the black hole after the evaporation process. Here, the remnant is formed at small scales and has a radius equal to the extremal value of the event horizon, enclosing a central singularity. This differs from the remnants of regular black holes, which do not enclose a singularity but instead typically surround a de Sitter core.

\section{DISCUSSION AND CONCLUSION}

The usual procedure to construct spherically symmetric (SS) black hole solutions sourced by matter consists of prescribing an energy density profile. For solutions satisfying $g_{tt} = -g_{rr}^{-1}$, this approach implicitly has an equation of state (EoS) $\rho = -p_r$. However, in Hořava gravity (HG) and Einstein–-{\ae}ther (EA) theory, for the SS case with a static {\ae}ther, this EoS is no longer satisfied. In this work, we have provided a procedure to obtain black hole solutions and to study their thermodynamic properties in HG--EA theory for the SS case with a static aether. This procedure consists of first specifying the form of the EoS, instead of adopting the aforementioned approach. In particular, we have investigated three cases in which the EoS is linear.

In the first case of study, we analyze the EoS $p_{r}(r) + p_{\theta}(r) = 0$, which leads to a solution that resembles the Reissner–Nordström black hole. The matter sources give rise to an electric-like potential but differ in their structure from the usual electromagnetic sources, whether those arising from the Maxwell electromagnetic tensor or from standard nonlinear electrodynamics. Thus, the source associated with this EoS can be viewed as an exotic anisotropic matter distribution induced by the modified gravitational dynamics of the HG--EA terms, which give rise to an effective electric–potential term in the geometry.

The second case of study corresponds to the EoS $\rho(r) = -p_{r}(r) - N \cdot p_{\theta}(r)$. The parameter $N$ can be regarded as an extension of the usual EoS $\rho = -p_r$, introduced by the HG--EA parameters. We obtain a non trivial solution such that the value of the event horizon is $n_{\text{odd}}$-fold degenerate. It is worth noting that the degeneracy of the event horizon leads to the temperature vanishing for $n_{\text{odd}} > 1$, with $n_{\text{odd}} \in {3,5,7,\ldots}$. Therefore, we are dealing with an extremal black hole. Moreover, following Wald’s procedure, we have found that, despite the vanishing temperature, the black hole possesses a non-zero entropy. In addition, the entropy obeys the area law. This is a non-trivial result, since, when other methodologies are employed, the entropy of black holes in the presence of matter usually does not follow the area law. The fact that $T_H = 0$ while the entropy is non-zero could be associated with the idea that the entropy depends only on the number of quantum states of the system.

In Case III, we study an EoS that represents an ultrarelativistic stiff fluid. We have obtained an asymptotically flat solution that can be viewed as the Schwarzschild metric plus a non trivial repulsive potential $C_{2}/r^{4\xi/\eta}$. In addition, this solution possesses both an inner horizon and an event horizon. The entropy also obeys the area law. The temperature exhibits two contributions. The first term has a negative slope, resembling the Schwarzschild temperature, which increases without bound as the mass and the horizon radius decrease, i.e. $T^{\text{Schw}}_H(M, r_h \to 0) \to \infty$. This term becomes dominant for large values of the event horizon. However, we note that the second term has a positive slope and becomes dominant at small scales. This behavior is consistent with the fact that the HG--EA terms are influential at short scales. We observe that the correction to the temperature at small scales, arising from the presence of the HG--EA terms, prevents the temperature from diverging to infinity as in the Schwarzschild case. In this way, the fact that the slope becomes positive at short scales causes the temperature to start decreasing after reaching a maximum, while approaching the value $T_H = 0$. In this situation, a black hole remnant is formed, which can be interpreted as what remains of the black hole after the evaporation process. Here, the remnant is formed at small scales and has a radius equal to the extremal value of the event horizon, enclosing a central singularity. This differs from the remnants of regular black holes, which do not enclose a singularity but instead typically surround a de Sitter core. We have displayed the behavior of the heat capacity. We note that a phase transition occurs between the unstable branch at large scales ($C < 0$) and the stable branch at short scales ($C > 0$), taking place at the same location where the temperature reaches its maximum.

\bibliography{mybib.bib}

\end{document}